# The Effect of Flow and Magnetic Twist on Resonant Absorption of Slow MHD Waves in Magnetic Flux Tubes

Mohammad Sadeghi[1,2], Karam Bahari[3], and Kayoomars Karami[1]
[1] Department of Physics, University of Kurdistan, Pasdaran Street, P.O. Box 66177-15175, Sanandaj, Iran
[2] Centre for Mathematical Plasma Astrophysics (CmPA), KU Leuven, Celestijnenlaan 200B bus 2400, B-3001 Leuven, Belgium
[3] Physics Department, Faculty of Science, Razi University, Kermanshah, Iran
*Received 2022 July 25; revised 2023 January 18; accepted 2023 January 19; published 2023 February 27*

## Abstract

Observations show that there are twisted magnetic flux tubes and plasma flow throughout the solar atmosphere. The main purpose of this work is to obtain the damping rate of sausage modes in the presence of magnetic twist and plasma flow. We obtain the dispersion relation for sausage modes in slow continuity in an inhomogeneous layer under the conditions of magnetic pores, then we solve it numerically. For the selected density profile, the magnetic field, and the plasma flow as a function of radius across the inhomogeneous layer, we show that the effect of the twisted magnetic field on the resonance absorption at low speed of the plasma flow is greater than one at high speed.

*Unified Astronomy Thesaurus concepts:* Solar physics (1476); Solar photosphere (1518); Solar atmosphere (1477)

## 1. Introduction

One of the proposed mechanisms for heating the solar corona is the propagation of magnetohydrodynamic (MHD) waves and their damping. The first observations of the transverse oscillations of the coronal loops were reported by Aschwanden et al. (1999) and Nakariakov et al. (1999). Nakariakov et al. (1999) reported oscillations with a period of 4.27 minutes and a damping time of 14.5 minutes. Wang & Solanki (2004) reported oscillations in the period of 3.9 minutes and damping times of 11.9 minutes using TRACE observations. The study of the MHD waves in the solar atmosphere is very important for solar seismography (e.g., Nakariakov & Verwichte 2005; De Moortel & Nakariakov 2012; Nakariakov et al. 2016) and atmospheric heating (e.g., Arregui 2015; De Moortel & Browning 2015), while the focus continues to be studying oscillation in the solar corona (Aschwanden et al. 1999; Nakariakov et al. 1999; Wang et al. 2002; Erdélyi & Taroyan 2008; Aschwanden & Schrijver 2011; Pascoe et al. 2016). Various modes of MHD waves in the lower layers of the solar atmosphere, in particular, the magnetic pores in the photosphere, have been identified and studied (e.g., Dorotovič et al. 2008, 2014; Jess et al. 2009, 2017; Morton et al. 2011; Tian et al. 2014; Grant et al. 2015; Moreels et al. 2015; Freij et al. 2016; Wang et al. 2018). Sausage waves are often studied in the magnetic pores in the Sun's atmosphere. Magnetic pores are relatively small (about 1–6 Mm in diameter) and have a field strength of about 1 kG, see, e.g., Sobotka (2003). The first observation of sausage modes in the large-scale loop oscillations in the corona was made by Aschwanden et al. (2004), while Dorotovič et al. (2008) reported the first observation of sausage waves in the lower layers of the solar atmosphere with a period of 20–70 minutes using the Swedish Solar Telescope. Dorotovič et al. (2014) studied magnetic pores and sunspots to find sausage modes and reported both slow and fast modes with a period of 4–65 minutes. They showed that sausage modes with many different wavelengths are present in photospheric structures. Grant et al. (2015) observed sausage waves in magnetic pores with a period of 181–412 s. They also showed that in magnetic pore conditions, the sausage wave has such strong damping that it disappears within 0.25 s of an oscillation period. Keys et al. (2018) reported that in magnetic pores, surface modes are more numerous than body modes, and that surface states carry more energy than body modes. They also found a period of 83–500 s for body modes and showed that for surface modes the period of oscillation is less than 100 s.

The resonant absorption of MHD waves to justify their damping was first introduced by Ionson (1978) in solar physics. In this mechanism, the energy of the global mode oscillations is transferred to the local Alfvén (cusp) perturbations in an inhomogeneous layer at the magnetic flux tube boundary, see, e.g., Goossens et al. (2011). While the resonance absorption under coronal conditions has been studied by many researchers (e.g., Ofman & Davila 1995; Goossens et al. 2002; Ruderman & Roberts 2002; Aschwanden et al. 2003; Terradas et al. 2006a, 2006b; Ruderman & Erdélyi 2009; Soler et al. 2013; Scherrer & McKenzie 2017), the role of resonance absorption in the lower atmosphere has not yet been fully studied (Bogdan et al. 1996; Ruderman 2009; Giagkiozis et al. 2016; Yu et al. 2017a, 2017b). In the lower atmosphere, in addition to the Alfvén resonance absorption, the cusp resonance absorption can be important. Yu et al. (2017a, 2017b) showed that for damping of slow surface kink and sausage waves in the photosphere, slow (cusp) resonance absorption can be a more effective mechanism than the Alfvén resonance absorption.

So far, many studies have been done on the resonance absorption in the presence of magnetic twist (Bennett et al. 1999; Erdélyi & Carter 2006; Erdélyi & Fedun 2006, 2007; Karami & Bahari 2010; Ebrahimi & Karami 2016; Giagkiozis et al. 2016; Sadeghi & Karami 2019) or plasma flow (Joarder et al. 1997; Soler et al. 2011; Bahari 2018; Sadeghi et al. 2021). Ruderman (2007) showed that the presence of a magnetic twist just inside the flux tube has no effect on the oscillation frequency of kink waves. Karami & Bahari (2010) investigated the effect of a twisted magnetic field on resonant absorption. They showed that for corona loops the period ratio $P_1/P_2$ of the fundamental mode to the first overtone mode of standing kink waves is less than 2. Karami & Bahari (2012) reported that if magnetic twist is considered in the annulus layer of the flux tube, it can affect the oscillation properties of kink waves. Giagkiozis et al. (2015) studied how the oscillation frequency of the fast sausage body waves is affected in the presence of a twisted magnetic field. Also, Giagkiozis et al. (2016) showed that the damping rate of sausage waves increases with increasing magnetic twist. Sadeghi & Karami (2019) investigated the slow resonance absorption for slow sausage surface waves in the presence of weak magnetic twist under magnetic pore conditions. They concluded that magnetic twist could increase the damping rate of the waves. Observations show that plasma flow in the magnetic flux tubes is present everywhere in the Sun's atmosphere, see, e.g., Brekke et al. (1997), Winebarger et al. (2001, 2002), Teriaca et al. (2004), Doyle et al. (2006), Ofman & Wang (2008), and Tian et al. (2008, 2009). Grant et al. (2015) showed that the average upflow speed in magnetic pores is about one-third the Alfvén speed, although they observed higher speed, up to about 1.15 Alfvén speed. Ruderman & Petrukhin (2019) studied the effect of flow on kink waves. They reported that the effect of flow on corona seismography is weak but has a significant effect on prominences. Bahari & Jahan (2020) investigated the effect of a twisted magnetic field and plasma flow on magnetic flux tubes. They showed that for a weak twist, the frequency of both the fundamental and first overtone modes decreases with increasing plasma flow. They also concluded that when both magnetic twist and plasma flow are considered in the flux tube the period ratio $P_1/P_2$ of the fundamental mode to the first overtone mode of standing kink waves is less than 2. Sadeghi et al. (2021) showed that the presence of background plasma flow can significantly increase the damping rate for slow sausage (kink) surface waves under magnetic pore conditions.

In this paper, our aim is to extend the studies conducted by Sadeghi & Karami (2019) and Sadeghi et al. (2021) to investigate the effect of the simultaneous presence of magnetic twist and plasma flow on the damping of slow sausage surface waves. In Section 2, the model and the basic equations of motion governing the MHD waves are presented. In Section 3, we find the dispersion relation in the case of no inhomogeneous layer and in the presence of an inhomogeneous layer. Our numerical results are discussed in Section 4. Finally, we conclude the paper in Section 5.

## 2. Model and Equations of Motion

### 2.1. Model of the Flux Tube

The linearized ideal MHD equations governing perturbations in magnetic flux tubes are given by Kadomtsev (1966) as

$$\rho\left(\frac{\partial}{\partial t} + \boldsymbol{V}\cdot\boldsymbol{\nabla}\right)^2 \boldsymbol{\xi} = -\boldsymbol{\nabla}\delta p - \frac{1}{\mu_0}(\delta\boldsymbol{B}\times(\boldsymbol{\nabla}\times\boldsymbol{B}) + \boldsymbol{B}\times(\boldsymbol{\nabla}\times\delta\boldsymbol{B})), \quad (1a)$$

$$\delta p = -\boldsymbol{\xi}\cdot\boldsymbol{\nabla}p - \gamma p\boldsymbol{\nabla}\cdot\boldsymbol{\xi}, \quad (1b)$$

$$\delta\boldsymbol{B} = -\boldsymbol{\nabla}\times(\boldsymbol{B}\times\boldsymbol{\xi}), \quad (1c)$$

where $\rho$, $p$, and $\boldsymbol{B}$ are the background density, kinetic pressure, and magnetic field, respectively. Also, $\boldsymbol{\xi}$ is the Lagrangian displacement vector, and $\delta p$ and $\delta\boldsymbol{B}$ are the Eulerian perturbations of the pressure and magnetic field, respectively. Here, $\gamma$ is the ratio of specific heats (taken to be 5/3 in this work), and $\mu_0$ is the magnetic permeability of free space.

We consider a stationary cylindrical flux tube model with a twisted magnetic field. The background magnetic field and plasma flow in cylindrical coordinates are defined as

$$\boldsymbol{B} = (0, B_\phi(r), B_z(r)), \quad (2)$$

$$\boldsymbol{V} = (0, V_\phi(r), V_z(r)). \quad (3)$$

The magnetohydrostatic equilibrium equation in the $r$ direction can be obtained from the zeroth-order ideal MHD equation and is given as

$$\frac{d}{dr}\left(p + \frac{B_\phi^2(r) + B_z^2(r)}{2\mu_0}\right) = \frac{\rho V_\phi^2(r)}{r} - \frac{B_\phi^2(r)}{\mu_0 r}. \quad (4)$$

We will use this equation to determine the dependence of the equilibrium plasma pressure on the radial coordinate. According to the model studied by Sadeghi & Karami (2019) and Sadeghi et al. (2021), we define the density profile, the magnetic field components, and the longitudinal component of the plasma flow as follows:

$$\rho(r) = \begin{cases} \rho_i, & r \leqslant r_i, \\ \rho_i\left(1 + (\zeta_\rho - 1)\left(\frac{r - r_i}{r_e - r_i}\right)\right), & r_i < r < r_e, \\ \rho_i\zeta_\rho, & r \geqslant r_e, \end{cases} \quad (5)$$

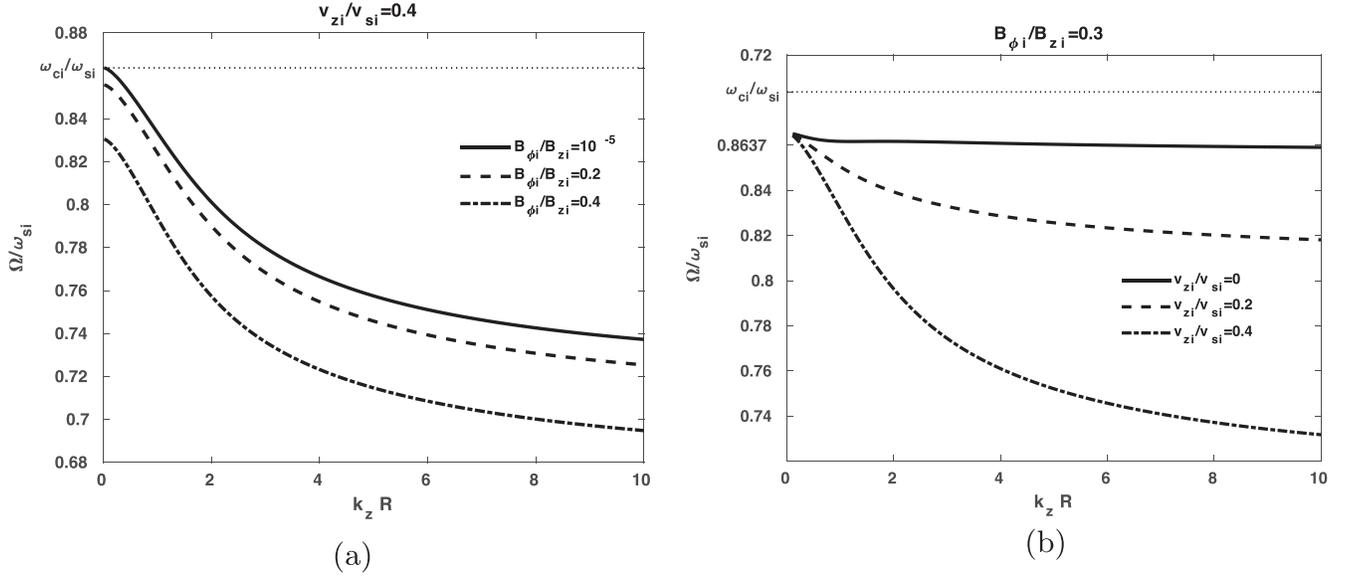

**Figure 1.** The Doppler shifted phase speed $\Omega/\omega_{si}$ obtained from Equation (40), for the slow surface sausage modes vs. $k_z R$ for various values of (a) the twist parameter $B_{\phi i}/B_{zi}$ and (b) the flow parameter $v_{zi}/v_{si}$. Under the magnetic pore conditions, following Grant et al. (2015) the auxiliary parameters are taken as $v_{Ai} = 12$ km s$^{-1}$, $v_{Ae} = 0$ km s$^{-1}$ (i.e., $B_{ze} = 0$), $v_{ze} = 0$ km s$^{-1}$, $v_{si} = 7$ km s$^{-1}$, $v_{se} = 11.5$ km s$^{-1}$, $v_{ci} = 6.0464$ km s$^{-1}$ ($\simeq 0.8638\, v_{si}$), and $v_{ce} = 0$ km s$^{-1}$.

$$B_z(r) = \begin{cases} B_0, & r \leqslant r_i, \\ B_0\sqrt{1 + (\zeta_{B_z}^2 - 1)\left(\dfrac{r - r_i}{r_e - r_i}\right)}, & r_i < r < r_e, \\ B_0 \zeta_{B_z}, & r \geqslant r_e, \end{cases} \quad (6)$$

$$B_\phi(r) = \begin{cases} Ar, & r \leqslant r_i, \\ Ar_i\sqrt{1 + (\zeta_{B_\phi}^2 - 1)\left(\dfrac{r - r_i}{r_e - r_i}\right)}, & r_i < r < r_e, \\ \zeta_{B_\phi} Ar_i, & r \geqslant r_e, \end{cases} \quad (7)$$

$$V_z(r) = \begin{cases} V_{zi}, & r \leqslant r_i, \\ V_{zi}\left(1 + (\zeta_{V_z} - 1)\left(\dfrac{r - r_i}{r_e - r_i}\right)\right), & r_i < r < r_e, \\ \zeta_{V_z} V_{zi}, & r \geqslant r_e. \end{cases} \quad (8)$$

In these equations $\rho_i$, $B_0$, and $V_{zi}$ are the plasma density, longitudinal magnetic field, and longitudinal plasma flow in the internal region of the tube, and all of them are constant quantities. The twist parameter $A$, which controls the azimuthal component of the magnetic field, is constant too. The quantities $\zeta_q$ are the ratio of the values of the quantities $q$ in the external and internal regions, where $q$ stands for the plasma density and the components of the background magnetic field plasma flow. For simplicity, we assume that the plasma flow is parallel to the background magnetic field. Using this condition, the azimuthal component of plasma flow can be obtained from Equations (6)–(8) as

$$V_\phi(r) = \frac{B_\phi}{B_z} V_z = \begin{cases} A_V r, & r \leqslant r_i, \\ A_V r_i [1 + \delta(\zeta_{V_z} - 1)]\sqrt{\dfrac{[1 + \delta(\zeta_{B_\phi}^2 - 1)]}{[1 + \delta(\zeta_{B_z}^2 - 1)]}}, & r_i < r < r_e, \\ \zeta_{V_\phi} A_V r_i, & r \geqslant r_e. \end{cases} \quad (9)$$

The parameters $A_V = \dfrac{A V_{zi}}{B_0}$ and $\zeta_{V_\phi} = \dfrac{\zeta_{B_\phi} \zeta_{V_z}}{\zeta_{B_z}}$, which give the values of the azimuthal plasma flow and its ratio in the external and internal regions, are not independent variables, and using the condition that the plasma flow and magnetic field be parallel, these quantities can be determined in terms of the previously defined parameters. Inserting Equations (5), (6), (7), and (9) into the magnetohydrostatic Equation (4), the background gas pressure is obtained as

$$p(r) = \begin{cases} p_0 - \left(1 - \dfrac{\mu_0 \rho_i A_V^2}{2A^2}\right)\dfrac{A^2 r^2}{\mu_0}, & r \leqslant r_i, \\ A_1 + A_2 r + A_3 r^2 + A_4 r^3 + A_5 \ln(r/r_i) + A_6 \ln\left(\dfrac{l + (r - r_i)(\zeta_{B_z}^2 - 1)}{l}\right), & r_i < r < r_e, \\ p_1 - \left(1 - \dfrac{\mu_0 \rho_i \zeta_\rho \zeta_{V_\phi}^2 A_V^2}{\zeta_{B_\phi}^2 A^2}\right)\dfrac{\zeta_{B_\phi}^2 A^2 r_i^2}{\mu_0}\ln(r/r_e), & r \geqslant r_e. \end{cases} \qquad (10)$$

In this equation, $p_0$ is a constant of integration, which gives the plasma pressure at the tube axis, and the other parameters are given in Appendix A.

The stationary model here is adjusted such that in the absence of plasma flow, it reduces to the equilibrium model considered by Sadeghi & Karami (2019), and in the absence of the magnetic twist, it recovers the model considered by Sadeghi et al. (2021).

In addition, Alfvén, sound, and cusp speeds are defined as follows:

$$v_{Ai}^2 \equiv \dfrac{B^2(r_i)}{\mu_0 \rho_i}, \qquad v_{Ae}^2 \equiv \dfrac{B^2(r_e)}{\mu_0 \rho_e}, \qquad (11)$$

$$v_{si}^2 \equiv \gamma \dfrac{p(r_i)}{\rho_i} = \gamma\left(p_0 - \left(1 - \dfrac{\mu_0 \rho_i A'^2}{2A^2}\right)A^2 r_i^2/\mu_0\right)/\rho_i, \quad v_{se}^2 \equiv \gamma \dfrac{p(r_e)}{\rho_e} = \gamma \dfrac{p_1}{\rho_e}, \qquad (12)$$

$$v_{c(i,e)}^2 \equiv \dfrac{v_{s(i,e)}^2 v_{A(i,e)}^2}{v_{s(i,e)}^2 + v_{A(i,e)}^2}. \qquad (13)$$

Here, $B^2 = B_\phi^2 + B_z^2$, and $v_{A(i,e)}$, $v_{s(i,e)}$, and $v_{c(i,e)}$ are the Alfvén, sound, and cusp speeds, and the subscripts $i$ and $e$ represent the internal and external regions, respectively. The $\beta$ parameter, i.e., the ratio of plasma pressure to magnetic pressure inside the flux tube is

$$\beta \equiv \dfrac{p(r_i)}{B^2(r_i)/(2\mu_0)} = 2v_{si}^2/(\gamma v_{Ai}^2). \qquad (14)$$

Using this equation, the integration constant $p_0$ can be determined in terms of the $\beta$ parameter as

$$\dfrac{p_0}{B_i^2/(2\mu_0)} = \beta + 2\left(1 - \dfrac{\mu_0 \rho_i A_V^2}{2A^2}\right)A^2 r_i^2/B_i^2, \qquad (15)$$

in which $B_i = B(r_i)$. Next, using Equation (12), which gives the speed of sound inside and outside the flux tube, and in Equation (14), we determine the ratio of densities of the external and internal regions as

$$\zeta_\rho \equiv \dfrac{\rho_e}{\rho_i} = \left(\dfrac{v_{si}}{v_{se}}\right)^2 \dfrac{p(r_e)}{\left[p_0 - \left(1 - \dfrac{\mu_0 \rho_i A_V^2}{2A^2}\right)A^2 r_i^2/\mu_0\right]},$$

$$= \dfrac{2\mu_0}{B_i^2 \beta}\left(\dfrac{v_{si}}{v_{se}}\right)^2 [B_i^2 \beta/2\mu_0 + A_2(r_e - r_i) + A_3(r_e^2 - r_i^2) + A_4(r_e^3 - r_i^3) + A_5 \ln(r_e/r_i), + A_6 \ln(1 + (\zeta_{B_z}^2 - 1))]. \qquad (16)$$

### 2.2. Governing Equations

Here, we use Equations 1(a)–(c) to obtain a differential equation for the radial displacement of the plasma, and then determine the solutions of the differential equation in various regions of the tube. Since the stationary quantities of the tube are time independent and depend only on the radial coordinate $r$, the perturbed quantities in cylindrical coordinates can be Fourier analyzed as

$$(\boldsymbol{\xi}, \delta P_T) \propto e^{i(m\phi + k_z z - \omega t)}, \qquad (17)$$

where $\omega$ is the angular frequency, $m$ is the azimuthal wavenumber for which only integer values are allowed, and $k_z$, is the longitudinal wavenumber. Also, $\delta P_T = \delta p + \boldsymbol{B} \cdot \delta \boldsymbol{B}/\mu_0$ is the Eulerian perturbation of total perturbed pressure. Using Equation (17) in

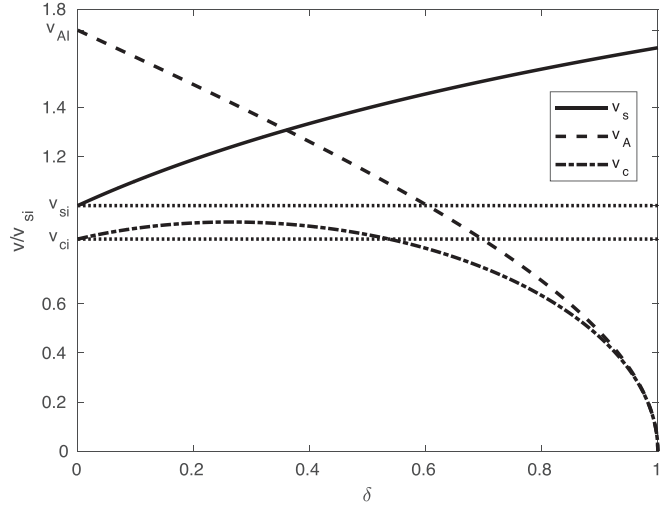

**Figure 2.** The sound, Alfvén, and cusp speeds as functions of $\delta$ representing radial coordinates in the inhomogeneous layer. The upper and lower horizontal dotted lines represent $v_{si}$ and $v_{ci}$, respectively, both normalized to the internal speed of sound. The parameters of the tube are $l/R = 0.1$, $B_{\phi i}/B_{zi} = 0.3$, $v_{Ai} = 12$ km s$^{-1}$, $v_{Ae} = 0$ km s$^{-1}$ (i.e., $B_{ze} = 0$), $v_{ze} = 0$ km s$^{-1}$, $v_{ze} = 0$ km s$^{-1}$, $v_{si} = 7$ km s$^{-1}$, $v_{se} = 11.5$ km s$^{-1}$, $v_{ci} = 6.0464$ km s$^{-1}$ ($\simeq 0.8638\, v_{si}$) and $v_{ce} = 0$ km s$^{-1}$.

the set of Equations 1(a)–(c), the following two first-order differential equations can be obtained:

$$\mathcal{D}\frac{d(r\xi)}{dr} = C_1(r\xi) - rC_2 \delta P_T, \tag{18a}$$

$$\mathcal{D}\frac{d\delta P_T}{dr} = \frac{1}{r}C_3(r\xi) - C_1 \delta P_T. \tag{18b}$$

These equations have been derived by Bondeson et al. (1987) and Goossens et al. (1992). The coefficients in Equations 18(a) and (b) are defined as

$$\mathcal{D} \equiv \rho(\Omega^2 - \omega_A^2)(v_s^2 + v_A^2)(\Omega^2 - \omega_c^2), \tag{19a}$$

$$C_1 \equiv Q\Omega^2 - 2m(v_s^2 + v_A^2)(\Omega^2 - \omega_c^2)T/r^2, \tag{19b}$$

$$C_2 \equiv \Omega^4 - \left(k_z^2 + \frac{m^2}{r^2}\right)(v_s^2 + v_A^2)(\Omega^2 - \omega_c^2), \tag{19c}$$

$$C_3 \equiv \rho\mathcal{D}\left[\Omega^2 - \omega_A^2 + \frac{2B_\phi}{\mu_0 \rho}\frac{d}{dr}\left(\frac{B_\phi}{r}\right) - 2v_\phi\frac{d}{dr}\left(\frac{v_\phi}{r}\right)\right] + Q^2 - 4(v_s^2 + v_A^2)(\Omega^2 - \omega_c^2)T^2/r^2, \tag{19d}$$

$$\Omega \equiv \omega - \omega_f, \qquad \omega_f \equiv \frac{m}{r}v_\phi + k_z v_z, \tag{19e}$$

$$Q \equiv -(\Omega^2 - \omega_A^2)\frac{\rho v_\phi^2}{r} + \frac{2\Omega^2 B_\phi^2}{\mu_0 r} + \frac{2\Omega f_B B_\phi v_\phi}{\mu_0 r}, \tag{19f}$$

$$T \equiv \frac{f_B B_\phi}{\mu_0} + \rho\Omega v_\phi, \tag{19g}$$

and

$$f_B \equiv \frac{m}{r}B_\phi + k_z B_z, \qquad \omega_A^2 \equiv \frac{f_B^2}{\mu_0 \rho}, \qquad \omega_c^2 \equiv \left(\frac{v_s^2}{v_A^2 + v_s^2}\right)\omega_A^2.$$

Combining the two first-order Equations 18(a) and (b), we obtain a second-order ordinary differential equation for the radial displacement, which has been determined by Goossens et al. (1992) as

$$\frac{d}{dr}\left[\frac{\mathcal{D}}{rC_2}\frac{d}{dr}(r\xi_r)\right] + \left[\frac{1}{\mathcal{D}}\left(C_3 - \frac{C_1^2}{C_2}\right) - r\frac{d}{dr}\left(\frac{C_1}{rC_2}\right)\right]\xi_r = 0. \tag{20}$$

Resonant absorption occurs under the condition $\mathcal{D} = 0$, which gives four Doppler shifted continua; two shifted Alfvén continua $\omega = \omega_f(r) \pm \omega_A(r)$, and two shifted slow continua $\omega = \omega_f(r) \pm \omega_c(r)$. Here, we study the slow surface waves propagating in the direction of the plasma flow. For these waves, in the model we consider here for the tube, both the conditions $\omega = \omega_f(r) + \omega_c(r)$ and $\omega = \omega_f(r) + \omega_A(r)$ are fulfilled. Yu et al. (2017b) showed that Alfvén resonant absorption is not an efficient damping mechanism for the slow surface kink waves. We generalize this result to the slow surface sausage waves and consider the slow resonant absorption as the only damping mechanism of the slow surface sausage waves. We postpone the investigation of Alfvén resonant absorption and validity of this generalization for a future work. If there is no plasma flow in the transitional layer, as in the case considered by Bahari et al. (2020) for the kink waves, it is possible that with increasing the flow speed sufficiently no further resonant absorption occurs and the MHD waves propagate without damping, but since we consider the plasma flow in the transitional layer, the condition for resonant absorption is fulfilled even for large flow speeds.

### 2.3. Solution Inside the Flux Tube

In this section, we obtain the solution of Equation (20) in the internal region of the tube. To do this we first simplify the coefficients given by Equation (19) for the sausage mode ($m = 0$) using the internal parameters of the tube

$$\mathcal{D}_i \equiv \rho_i(\Omega_i^2 - \omega_{Ai}^2), \tag{21a}$$

$$C_1 \equiv \left(\frac{2n_i^2 A^2}{\mu_0} + \frac{2n_i^2 B_{zi} A A_V}{\Omega_i \mu_0} - \rho_i n_i^2 (\omega_{Ai}^2 - \Omega_i^2) A_V^2\right)r, \tag{21b}$$

$$C_2 \equiv n_i^2 - k_z^2, \tag{21c}$$

$$C_3 \equiv \rho_i^2\left((\Omega_i^2 - \omega_{Ai}^2)^2 + Q_i^2 - 4\left(\frac{\omega_{Ai}^2 A^2}{\mu_0 \rho_i} + \Omega_i^2 A_V^2 + \frac{2\Omega_i \omega_{Ai} A A_V}{\mu_0^{1/2} \rho_i^{1/2}}\right)\right), \tag{21d}$$

$$Q_i^2 \equiv \left(\frac{4n_i^2 A^4}{\mu_0^2 \rho_i^2} + n_i^2 A_V^4 - \frac{2n_i^2 \omega_{Ai}^2 A_V^4}{\Omega_i^2} + \frac{n_i^2 \omega_{Ai}^4 A_V^4}{\Omega_i^4} + \frac{4(1-\Omega_i^2)n_i^2 \omega_{Ai} A A_V^3}{\Omega_i^3 \mu_0^{1/2} \rho_i^{1/2}} + \frac{8n_i^2 \omega_{Ai} A^3 A_V}{\Omega_i \mu_0^{3/2} \rho_i^{3/2}} - \frac{4n_i^2 \omega_{Ai}^2 A^2 A_V^2}{\mu_0 \rho_i}\right)r^2,$$

$$\omega_{fi} \equiv k_z v_{zi}, \quad \Omega_i = \omega - \omega_{fi}, \tag{21e}$$

$$n_i^2 \equiv \frac{\Omega_i^4}{(v_{Ai}^2 + v_{si}^2)(\Omega_i^2 - \omega_{ci}^2)}. \tag{22}$$

With these simplifications the differential equation governing the radial displacement becomes

$$\mathcal{R}^2 \frac{d^2 \xi_r}{d\mathcal{R}^2} + \mathcal{R}\frac{d \xi_r}{d\mathcal{R}} - \left(1 + \frac{k_{ri}^2}{k_z^2}\mathcal{R}^2 + E\mathcal{R}^4\right)\xi_r = 0, \tag{23}$$

where

$$\mathcal{R} = k_\alpha r, \tag{24}$$

$$k_\alpha \equiv k_z(1-\alpha^2)^{1/2}, \tag{25}$$

$$\alpha^2 \equiv \frac{4A^2\omega_{Ai}^2}{\mu_0 \rho_i(\omega^2 - \omega_{Ai}^2)^2} + \frac{\Omega_i^2 A_V^2}{(\omega^2 - \omega_{Ai}^2)^2} + \frac{2\Omega_i \omega_{Ai} A A_V}{\mu_0^{1/2}\rho_i^{1/2}(\omega^2 - \omega_{Ai}^2)^2}, \tag{26}$$

$$E \equiv \frac{\rho_i^2 k_{ri}^2 Q_i^2 - C_1^2}{\mathcal{D}_i^2 k_z^2 (1-\alpha^2)^2 r^2}, \tag{27}$$

$$k_{ri}^2 \equiv -(n_i^2 - k_z^2)\left(=\frac{(\omega_{si}^2 - \Omega_i^2)(\omega_{Ai}^2 - \Omega_i^2)}{(v_{Ai}^2 + v_{si}^2)(\omega_{ci}^2 - \Omega_i^2)}\right). \tag{28}$$

Here, we have defined the frequencies $\omega_{s(i,e)} = kv_{s(i,e)}$. The solution of Equation (23) for the radial component of the Lagrangian displacement $\boldsymbol{\xi}$ inside the flux tube ($r \leqslant r_i$) for sausage modes ($m = 0$) is as follows (Giagkiozis et al. 2015, 2016):

$$\xi_{ri}(s) = A_i \frac{s^{1/2}}{E^{1/4}} e^{-s/2} M(a, b; s), \tag{29}$$

where $A_i$ is a constant and $M(.)$ is the Kummer function (Abramowitz & Stegun 2012). Using Equation (29) in Equation 18(a) the Eulerian perturbation of total pressure relation inside the flux tube can be obtained as

$$\delta P_{Ti}(s) = A_i e^{-s/2}\left(\frac{k_a \mathcal{D}_i}{n_i^2 - k_z^2}\right)\left[\left(\frac{n_i + k_z}{k_z}\right)sM(a, b; s) - 2M(a, b-1; s)\right], \tag{30}$$

in which

$$a \equiv 1 + \frac{k_{ri}^2}{4k_z^2 E^{1/2}}, \quad b = 2, \quad s \equiv k_\alpha^2 E^{1/2} r^2. \quad (31)$$

### 2.4. Solution Outside the Flux Tube

Now we determine the solution of Equation (20) in the external region of the tube. The coefficients given by Equation (19) for the sausage mode ($m = 0$) using the parameters of the tube in the external region are

$$\mathcal{D}_e \equiv \rho_e(\Omega_e^2 - \omega_{Ae}^2), \quad (32a)$$

$$C_1 \equiv \left(\frac{2n_e^2 \zeta_{B_\phi}^2 A^2 r_i^2}{\mu_0} + \frac{2n_e^2 B_0 A A_V \zeta_{B_z} \zeta_{B_\phi} \zeta_{V_\phi} r_i^2}{\Omega_e \mu_0} - \rho_e n_e^2 (\omega_{Ae}^2 - \Omega_e^2) A_V \zeta_{V_\phi} r_i \right) \frac{1}{r}, \quad (32b)$$

$$C_2 \equiv n_e^2 - k_z^2, \quad (32c)$$

$$C_3 \equiv \rho_e^2 \left( (\Omega_e^2 - \omega_{Ae}^2)^2 + Q^2 - 4\left(\frac{\omega_{Ae}^2 \zeta_{B_\phi}^2 A^2 r_i^2}{\mu_0 \rho_e} + \Omega_e^2 \zeta_{v_\phi}^2 A_v^2 r_i^2 + \frac{2\Omega_e \omega_{Ae} \zeta_{B_\phi} \zeta_{V_\phi} A A_V r_i^2 v_{\phi e}}{\mu_0^{1/2} \rho_e^{1/2}}\right) \frac{1}{r^2} \right), \quad (32d)$$

$$Q_e^2 \equiv \left(\frac{4n_e^2 \zeta_{B_\phi}^4 A^4 r_i^4}{\mu_0^2 \rho_e^2} + n_e^2 \zeta_{v_\phi}^4 A_v^4 r_i^4 - \frac{2n_e^2 \omega_{Ae}^2 \zeta_{v_\phi}^4 A_V^4 r_i^4}{\Omega_e^2} + \frac{n_e^2 \omega_{Ae}^4 \zeta_{V_\phi}^4 A_V^4 r_i^4}{\Omega_e^4}\right.$$
$$\left. + \frac{4(1 - \Omega_e^2) n_e^2 \omega_{Ae} \zeta_{B_\phi} \zeta_{V_\phi}^3 A A_V^3 r_i^4}{\Omega_e^3 \mu_0^{1/2} \rho_e^{1/2} r_e^4} + \frac{8n_e^2 \omega_{Ae} \zeta_{B_\phi}^3 \zeta_{V_\phi} A^3 A_V r_i^4}{\Omega_e \mu_0^{3/2} \rho_e^{3/2}} - \frac{4n_e^2 \omega_{Ae}^2 \zeta_{B_\phi}^2 \zeta_{V_\phi}^2 A^2 A_V^2 r_i^4}{\mu_0 \rho_e}\right) \frac{1}{r^2}, \quad (32e)$$

$$\omega_{fe} \equiv k_z v_{ze}, \quad \Omega_e = \omega - \omega_{fe}. \quad (32f)$$

Hence, the differential Equation (20) governing the radial displacement outside the flux tube becomes

$$r^2 \frac{d^2 \xi_r}{dr^2} + r\frac{d\, \xi_r}{dr} - (k_{re} r^2 + \nu^2)\xi_r = 0. \quad (33)$$

The solution of this equation with physically acceptable behavior, i.e., vanishing for large values of the radial coordinate is

$$\xi_{re}(r) = A_e K_\nu(k_{re} r). \quad (34)$$

Here $K(.)$ is the modified Bessel function of the second kind (Abramowitz & Stegun 2012). Substituting solution (34) in Equation 18(a), the Eulerian perturbation of total pressure outside the flux tube can be obtained as

$$\delta P_{Te}(r) = A_e \left[ \left(\frac{\mu_0 (1 - \nu)\mathcal{D}_e - 2\zeta_{B_\phi}^2 A^2 r_i^2 n_e^2}{\mu_0 r (k_z^2 - n_e^2)}\right) K_\nu(k_{re} r) - \frac{\mathcal{D}_e}{k_{re}} K_{\nu-1}(k_{re} r) \right]. \quad (35)$$

In Equations (34) and (35), the following definitions are used:

$$k_{re}^2 \equiv -(n_e^2 - k_z^2)\left(=\frac{(\omega_{se}^2 - \Omega_e^2)(\omega_{Ae}^2 - \Omega_e^2)}{(v_{Ae}^2 + v_{se}^2)(\omega_{ce}^2 - \Omega_e^2)}\right), \quad (36)$$

$$\nu^2(r) \equiv 1 + \frac{1}{\mathcal{D}_e^2}\left(\rho_e^2 k_{re}^2 Q_e^2 - 4k_{re}^2 \rho_e^2 \left(\frac{\omega_{Ae}^2 \zeta_{B_\phi}^2 A^2 r_i^2}{\mu_0 \rho_e} + \Omega_e^2 v_{\phi e}^2 + \frac{2\Omega_e \omega_{Ae} \zeta_{B_\phi} \zeta_{V_\phi} A A_V r_i^2}{\mu_0^{1/2} \rho_e^{1/2}}\right)\right.$$
$$+ \left(\frac{2n_e^2 \zeta_{B_\phi}^2 A^2 r_i^2}{\mu_0} + \frac{2n_e^2 B_0 A A_V \zeta_{B_z} \zeta_{B_\phi} \zeta_{V_\phi} r_i^2}{\Omega_e \mu_0} - \rho_e n_e^2 (\omega_{Ae}^2 - \Omega_e^2)\zeta_{V_\phi} A_V r_i\right)^2$$
$$\left. + \mathcal{D}_e \left(\frac{2n_e^2 \zeta_{B_\phi}^2 A^2 r_i^2}{\mu_0} + \frac{2n_e^2 B_0 A A_V \zeta_{B_z} \zeta_{B_\phi} \zeta_{V_\phi} r_i^2}{\Omega_e \mu_0} - \rho_e n_e^2 (\omega_{Ae}^2 - \Omega_e^2)\zeta_{V_\phi} A_V r_i\right)\right). \quad (37)$$

If $v_\phi = 0$, this equation reduces to

$$\nu^2(r) \equiv 1 + 2\frac{\zeta_{B_\phi}^2 A^2 r_i^2}{\mu_0^2 \mathcal{D}_e^2}(2\zeta_{B_\phi}^2 A^2 r_i^2 n_e^2 k_z^2 + \mu_0 \rho_e [\omega_{Ae}^2(3n_e^2 - k_z^2) - \omega^2(n_e^2 + k_z^2)]). \quad (38)$$

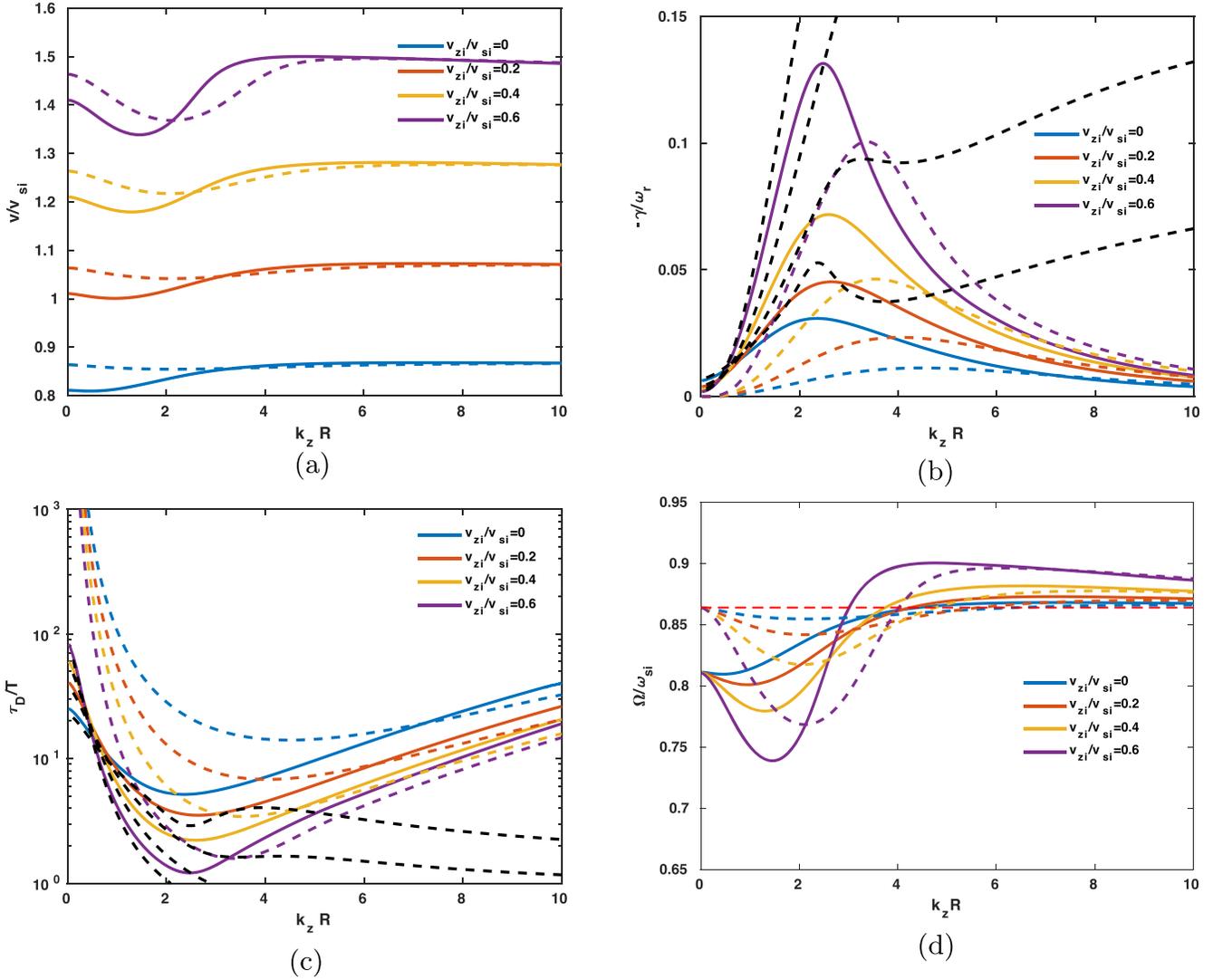

**Figure 3.** (a) The phase speed $v/v_{si} \equiv \omega_r/\omega_{si}$, (b) the damping rate to frequency ratio $-\gamma/\omega_r$, (c) the damping time to period ratio $\tau_D/T = \omega/(2\pi|\gamma|)$, and (d) the Doppler shifted phase speed $\Omega/\omega_{si}$ as functions of $k_zR$ for $l/R = 0.1$ of the slow surface sausage modes for various values of the flow parameter. Here, the dashed and solid curves correspond to the case of $B_{\phi i}/B_{zi} = 0$ and $B_{\phi i}/B_{zi} = 0.3$, respectively. The black dashed curves show the analytical results obtained from Equation (55). Other parameters of the tube are $v_{Ai} = 12$ km s$^{-1}$, $v_{Ae} = 0$ km s$^{-1}$ (i.e., $B_{ze} = 0$), $v_{ze} = 0$ km s$^{-1}$, $v_{si} = 7$ km s$^{-1}$, $v_{se} = 11.5$ km s$^{-1}$, $v_{ci} = 6.0464$ km s$^{-1}$ ($\simeq 0.8638\, v_{si}$), and $v_{ce} = 0$ km s$^{-1}$.

In the limit of vanishing flow speed, all the solutions obtained in the Sections 2.3 and 2.4 reduce to the corresponding solutions determined by Giagkiozis et al. (2015) and Sadeghi & Karami (2019), and in the limits of vanishing magnetic twist, the mentioned solutions reduce to the corresponding solutions determined by Sadeghi et al. (2021).

## 3. Dispersion Relation

### 3.1. Dispersion Relation for the Case of No Inhomogeneous Layer

Now, we obtain the dispersion relation for the sausage modes in the absence of the inhomogeneous layer using the appropriate boundary conditions at the flux tube boundary. The radial displacement and Lagrangian perturbation of the total pressure must be continuous at the tube boundary

$$\boldsymbol{\xi}_{ri}|_{r=R} = \boldsymbol{\xi}_{re}|_{r=R}, \tag{39a}$$

$$\left(\delta P_{Ti} - \frac{B_{\phi i}^2}{\mu_0 r}\boldsymbol{\xi}_{ri}\right)\bigg|_{r=R} = \left(\delta P_{Te} - \frac{B_{\phi e}^2}{\mu_0 r}\boldsymbol{\xi}_{re}\right)\bigg|_{r=R}, \tag{39b}$$

where $R$ is the tube radius. Inserting the solutions given by Equations (29), (29), (30), and (35) in the above boundary conditions, after some algebra the dispersion relation can be obtained as follows:

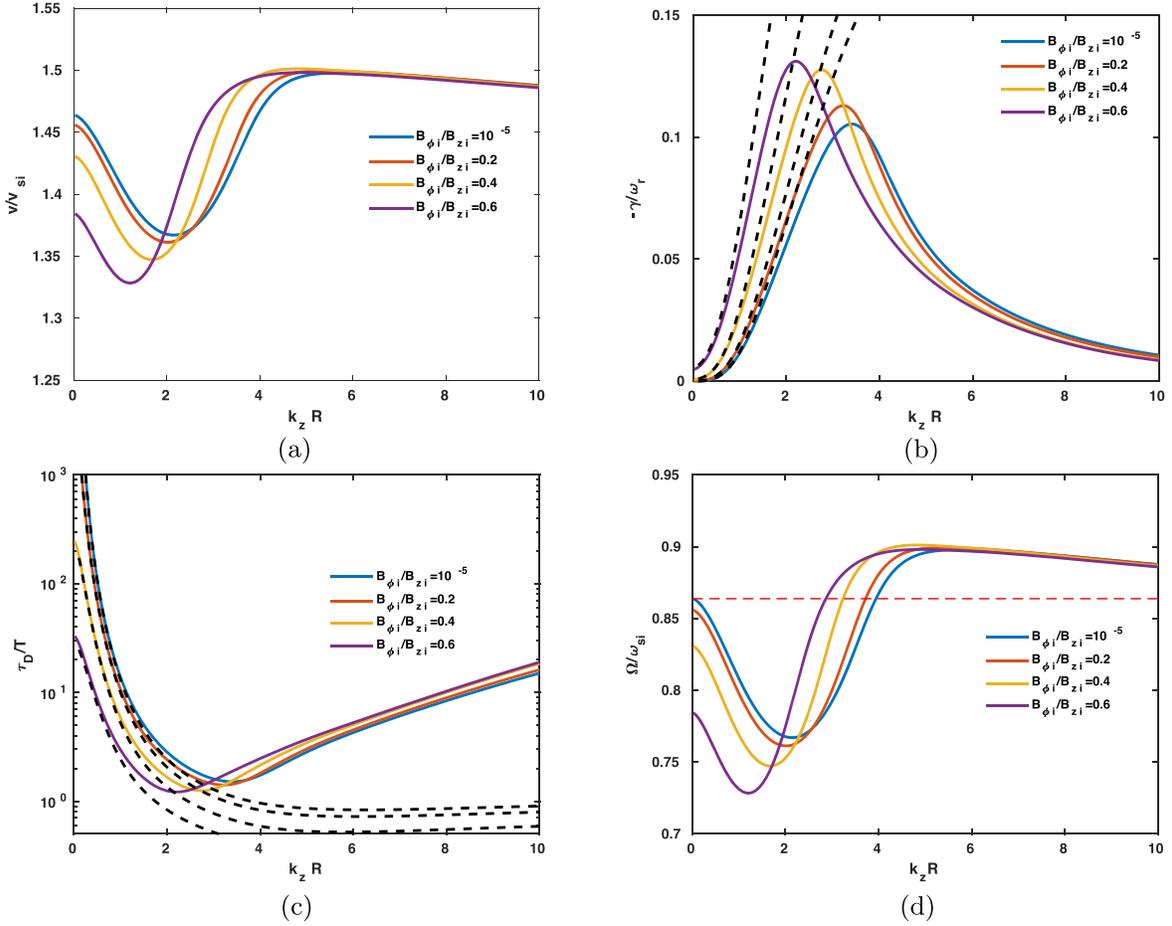

**Figure 4.** (a) The phase speed $v/v_{si} \equiv \omega_r/\omega_{si}$, (b) the damping rate to frequency ratio $-\gamma/\omega_r$, (c) the damping time to period ratio $\tau_D/T = \omega/(2\pi|\gamma|)$, and (d) the Doppler shifted phase speed $\Omega/\omega_{si}$ as functions of $k_z R$ for $l/R = 0.1$ and the flow parameter $v_{zi}/v_{si} = 0.6$ of the slow surface sausage modes for various values of the twist parameter. The black dashed curves show the analytical results obtained from Equation (55). Other parameters of the tube are the same as in Figure 3.

$$-\frac{\mu_0 D_i}{k_{ri}^2}\left[\left(\frac{n_i + k_z}{k_z}\right)s - 2\frac{M(a, b-1, s)}{M(a, b, s)}\right] = \frac{\mu_0 D_e}{k_{re}^2}\left(1 - \nu - \frac{k_{re}R\, K_{\nu-1}(k_{re}R)}{K_\nu(k_{re}R)}\right) - \left(1 + \frac{2n_e^2}{k_{re}^2}\right)\zeta_{B_\phi}^2 A^2 R^2 + A^2 R^2. \quad (40)$$

Here, we solve this dispersion relation numerically. In Figure 1, we plot the Doppler shifted phase speed $\Omega/\omega_{si}$ versus $k_z R$ of the slow surface sausage modes obtained from Equation (40). Panel (a) shows various values of the magnetic twist parameter and the flow parameter $v_{zi}/v_{si} = 0.4$. This panel shows that at a given flow the Doppler shifted phase speed decreases with increasing magnetic twist parameter. Also, for a small magnetic twist (i.e., $B_{\phi i}/B_{zi} = 10^{-5}$), the results are consistent with the results of Sadeghi et al. (2021). Panel (b) shows that with increasing flow parameter at a given magnetic twist ($B_{\phi i}/B_{zi} = 0.3$), the value of the Doppler shifted phase speed decreases and the rate of decrease is greater at larger $k_z R$. This panel shows for the case of no flow (i.e., $v_{zi}/v_{si} = 0$), the result of Sadeghi & Karami (2019) (dashed line) is recovered.

### 3.2. Dispersion Relation in the Presence of an Inhomogeneous Layer and Resonant Absorption

Following Sakurai et al. (1991), we do not need to solve the differential Equation (20) in the inhomogeneous layer. Based on the results of Sakurai et al. (1991), we can connect the solutions inside and outside the flux tube by the connection formula as follows:

$$[\xi_r] \equiv \xi_{re}(r_e) - \xi_{ri}(r_i), = \frac{-i\pi\mu\omega_c^4}{|\Delta_c|rB^2\omega_A^2}\bigg|_{r=r_c}\left(\delta p_{Ti} - \frac{Q_i\xi_{ri}}{\omega_c^2}\right)\bigg|_{r=r_i}, \quad (41a)$$

$$[\delta p_T] \equiv \delta p_{Te}(r_e) - \delta p_{Ti}(r_i), = \frac{-i2\pi\omega_c^2 Q}{|\Delta_c|rB^2\omega_A^2}\bigg|_{r=r_c}\left(\delta p_{Ti} - \frac{Q_i\xi_{ri}}{\omega_c^2}\right)\bigg|_{r=r_i}, \quad (41b)$$

where $[\xi_r]$ and $[\delta p_T]$ are the jump conditions in the Lagrangian radial displacement and total pressure perturbation, respectively, across the inhomogeneous (resonant) layer. The subscript $c$ denotes the position of the slow resonance ($r = r_c$) and $|\Delta_c| \equiv \left|\frac{d(\Omega^2 - \omega_c^2)}{dr}\right|_{r=r_c}$.

Inserting the solutions of Equation (20) inside Equations (29) and 30), and outside Equations (34) and (35), the flux tube in the connection formulae 41(a) and (b), the dispersion relation for the slow surface sausage waves in the presence of magnetic twist and plasma flow can be obtained as follows:

$$D_{AR} + i D_{AI} = 0, \quad (42)$$

in which $D_{AR}$ and $D_{AI}$ are given by

$$D_{AR} = \rho_i(\Omega_i^2 - \omega_{Ai}^2) - r_i \rho_e (\Omega_e^2 - \omega_{Ae}^2) \frac{k_{ri}}{k_{re}} \mathcal{Q}, \quad (43)$$

$$D_{AI} = \frac{\pi \rho_i \rho_e k_z^2}{k_{re} \rho_c |\Delta_c|}\Bigg|_{r=r_c} \left(\frac{v_{sc}^2}{v_{Ac}^2 + v_{sc}^2}\right)^2 \\
\times ((\Omega_i^2 - \omega_{Ai}^2) + \mathcal{Z})(\Omega_e^2 - \omega_{Ae}^2) \mathcal{G}, \quad (44)$$

where

$$\mathcal{Q} \equiv -k_{ri} r_i \frac{\left(\frac{\mu_0 D_e(1-\nu) - 2\zeta_{B_\phi} A r_i n_e^2}{D_e \mu_0 r_e k_{re}} - \frac{K_{\nu-1}(k_{re}r_e)}{K_\nu(k_{re}r_e)}\right)}{\left[\frac{n_i + k_z}{k_z}s - 2\frac{M(a, b-1; s)}{M(a, b; s)}\right]}, \quad (45)$$

$$\mathcal{G} \equiv \left(-\frac{\mathcal{Q}k_{re}}{\omega_c^2 D_e}\bigg|_{r=r_c} + \frac{\mu_0(1-\nu)D_e - 2\zeta_{B_\phi}^2 A^2 r_i^2 n_e^2}{\mu_0 r_e k_{re} D_e}\right. \\
\left. - \frac{K_{\nu-1}(k_{re}r_e)}{K_\nu(k_{re}r_e)}\right), \quad (46)$$

$$\mathcal{Z} \equiv \frac{Q_i^2 k_{ri}^2}{\omega_{ci}^2 \left[\frac{n_i + k_z}{k_z}s - 2\frac{M(a, b-1; s)}{M(a, b; s)}\right]}. \quad (47)$$

Using the equations that give the background values of the density, magnetic field, and plasma pressure, we define the quantities of sound, Alfvén, and cusp speed in the inhomogeneous layer ($r_i < r < r_e$) as

$$v_s^2 = v_{si}^2 \left[\frac{1 + \delta(\zeta_\rho v_{sei}^2 - 1) + \Lambda}{1 + \delta(\zeta_\rho - 1)}\right], \quad (48)$$

$$v_A^2 = v_{Ai}^2 \left[\frac{1 + \delta(\zeta_\rho v_{Aei}^2 - 1)}{1 + \delta(\zeta_\rho - 1)}\right], \quad (49)$$

$$v_c^2 = \frac{v_{si}^2 v_{Ai}^2}{[1 + \delta(\zeta_\rho - 1)]} \\
\times \frac{[1 + \delta(\zeta_\rho v_{sei}^2 - 1) + \Lambda][1 + \delta(\zeta_\rho v_{Aei}^2 - 1)]}{[v_{si}^2(1 + \delta(\zeta_\rho v_{sei}^2 - 1) + \Lambda) + v_{Ai}^2(1 + \delta(\zeta_\rho v_{Aei}^2 - 1))]}, \quad (50)$$

where $\delta \equiv \frac{r - r_i}{r_e - r_i}$, $\zeta_\rho \equiv \rho_e/\rho_i$, $v_{sei} \equiv v_{se}/v_{si}$, $v_{Aei} \equiv v_{Ae}/v_{Ai}$ and

$$\Lambda \equiv \frac{\gamma}{v_{si}^2}[A_3(r^2 - \delta r_e^2) + A_4(r^3 - \delta r_e^3) + A_5(\ln(r/r_i) - \delta \ln(r_e/r_i)) + A_6(\ln(1 + \delta(\zeta_{B_z}^2 - 1)) - \delta \ln(1 + (\zeta_{B_z}^2 - 1))]. \quad (51)$$

Under the photospheric conditions, the parameter $\Lambda$ is simplified as

$$\Lambda \equiv \frac{\gamma}{v_{si}^2}[A_3(r^2 - \delta r_e^2) + A_4(r^3 - \delta r_e^3) + A_5(\ln(r/r_i) - \delta \ln(r_e/r_i))]. \quad (52)$$

In Figure 2, we have plotted the background sound, Alfvén, and cusp speeds as a function of $\delta$ in the transitional layer. Because the Doppler shifted oscillation frequency, as is obvious in Figure 1, is smaller than $v_{si}$, then the conditions of both slow and Alfvén resonant damping are fulfilled. But as we discussed in Section 2.2 we ignore the Alfvén resonant absorption. Note that the cusp resonance occurs at a place where the Doppler shifted frequency equals the cusp frequency, i.e., $\Omega = \omega_c = k_z v_c$. Like Yu et al. (2017b) and Sadeghi & Karami (2019) the position of the cusp resonance point $r_c$ can be obtained by putting $v_c$ in Equation (50) equal to $\Omega/k_z$ and write it as an algebraic equation for $\delta_c \equiv \delta|_{r=r_c} = \frac{r_c - r_i}{r_e - r_i}$ as follows:

$$\mathcal{A}\delta_c^4 + \mathcal{B}\delta_c^3 + \mathcal{C}\delta_c^2 + \mathcal{E}\delta_c + \mathcal{F} = 0. \tag{53}$$

The constants $\mathcal{A}, \mathcal{B}, \mathcal{C}, \mathcal{E}$, and $\mathcal{F}$ are given in Appendix B. Equations (42) and (53) can be solved simultaneously and numerically to give the oscillation frequency, damping rate, and resonant point of the waves.

### 3.3. Weak Damping Limit—Slow Continuum

In the weak damping limit an analytical expression for the damping rate can be determined from the dispersion relation (42). Note that in Equations (43) and (44) we have the complex Doppler shifted frequency $\Omega = \omega - k_z v_z$ where $\omega = \omega_r + i\gamma$, in which $\omega_r$ and $\gamma$ are the cusp frequency and the damping rate, respectively. The damping rate $\gamma$ in the limit of weak damping, i.e., $\gamma \ll \omega_r$ is given as (Goossens et al. 1992)

$$\gamma = -D_{\mathrm{AI}}(\omega_r) \left( \frac{\partial D_{\mathrm{AR}}}{\partial \omega} \bigg|_{\omega_r} \right)^{-1}. \tag{54}$$

Using Equation (54), we obtain an analytical relation for the damping rate $\gamma$ (see Appendix C) for the slow surface sausage waves as follows:

$$\gamma = -\frac{\pi \rho_e k_z^2}{k_{re} \rho_c |\Delta_c|} \bigg|_{r=r_c} \left( \frac{v_s^2}{v_A^2 + v_s^2} \right)^2 \\
\times \frac{((\Omega_i^2 - \omega_{Ai}^2) + \mathcal{Z})(\Omega_e^2 - \omega_{Ae}^2)^2 \mathcal{G}}{2\left(\Omega_i - \chi \Omega_e \frac{k_{ri}}{k_{re}} \mathcal{Q}\right) - \Omega_e \chi \mathcal{T}}, \tag{55}$$

where the quantity $\mathcal{T}$ is given by Equation (C14). Note that in the limit of no twist, i.e., $B_{\phi i} = B_{\phi e} = 0$, Equations (45), (46), (47), (C7), and (C8) reduce to

$$\mathcal{Q} = \mathcal{Q}_0 \equiv \frac{I_0'(x) K_0(y)}{I_0(x) K_0'(y)},$$

$$\mathcal{G} = \mathcal{G}_0 \equiv \frac{K_0(y)}{K_0'(y)},$$

$$\mathcal{Z} = 0,$$

$$\mathcal{P} = \mathcal{P}_0 \equiv \left( \frac{I_0''(x)}{I_0(x)} - \frac{I_0'(x)^2}{I_0(x)^2} \right) \frac{K_0(y)}{K_0'(y)},$$

$$\mathcal{S} = \mathcal{S}_0 \equiv \left( 1 - \frac{K_0''(y) K_0(y)}{K_0'(y)^2} \right) \frac{I_0'(x)}{I_0(x)}, \tag{56}$$

where $x = k_{ri} r_i$ and $y = k_{re} r_e$, and we have also used the relation

$$\lim_{a \to \infty} \frac{M(a, b-1, s)}{M(a, b, s)} = \frac{x}{2} \frac{I_0(x)}{I_1(x)}. \tag{57}$$

In the case of a straight magnetic field, Equations (56) and (57) can be used to simplify $\mathcal{T}$ and consequently the damping rate given by Equation (55) reduces to Equation (40) in Sadeghi et al. (2021) for the slow sausage surface modes in the presence of flow.

## 4. Numerical Results

In this section, we obtain the frequency and damping rate of the slow surface sausage modes using the numerical solution of the dispersion relation (42) under the magnetic pore conditions. The period ratio $P_1/P_2$ is not a point of interest in the present paper because it is relevant in the case of standing waves but here we study the propagating waves. Following Grant et al. (2015), we set the flux tube parameters as $v_{Ai} = 12 \text{ km s}^{-1}$, $v_{Ae} = 0 \text{ km s}^{-1}$ (i.e., $B_{\phi e} = B_{ze} = 0$), $v_{si} = 7 \text{ km s}^{-1}$, $v_{se} = 11.5 \text{ km s}^{-1}$, $v_{ci} = 6.0464 \text{ km s}^{-1}$ ($\simeq 0.8638 \, v_{si}$), and $v_{ce} = 0 \text{ km s}^{-1}$. The numerical results are shown in Figures 3–7.

Figure 3 represents the variation of the phase speed (or normalized frequency) $v/v_{si} = \omega_r/k_z v_{si} \equiv \omega_r/\omega_{si}$, the damping rate to frequency ratio $-\gamma/\omega_r$, the damping time to period ratio $\tau_D/T = \omega/(2\pi|\gamma|)$, and the Doppler shifted phase speed $\Omega/\omega_{si}$ of the slow surface sausage modes versus $k_z R$ for various values of the flow parameter $v_{zi}/v_{si} = (0, 0.2, 0.4, 0.6)$ and for the thickness of the inhomogeneous layer $l/R = 0.1$. The dashed and solid curves in Figure 3, correspond to the cases $B_{\phi_i}/B_{zi} = 0$ and $B_{\phi_i}/B_{zi} = 0.3$, respectively. The black dashed curves show the results obtained from Equation (55) for the weak damping limit. The figure shows that (i) the value of the phase speed $v/v_{si}$ increases with increasing the flow parameter $v_{zi}/v_{si}$. (ii) As the flow speed increases, the maximum value of $-\gamma/\omega_r$ increases too. Also, in twisted flux tubes the minimum value of $-\gamma/\omega_r$ occurs at smaller $k_z R$ in comparison with untwisted tubes. (iii) The minimum value of the damping time to period ratio decreases when the flow speed increases. (iv) The effect of plasma flow in decreasing the ratio of the damping time to the oscillation period is more efficient in untwisted tubes; for example, the value of $\tau_D/T$ for $v_{zi}/v_{si} = 0.6$ and $B_{\phi_i}/B_{zi} = 0.5$ decreases $\sim 45\%$ but for $v_{zi}/v_{si} = 0.6$ and $B_{\phi_i}/B_{zi} = 0$ decreases $\sim 88\%$ in comparison with the case of the static tube. (v) The minimum value of Doppler shifted phase speed $\Omega/\omega_{si}$ decreases with increasing the flow parameter.

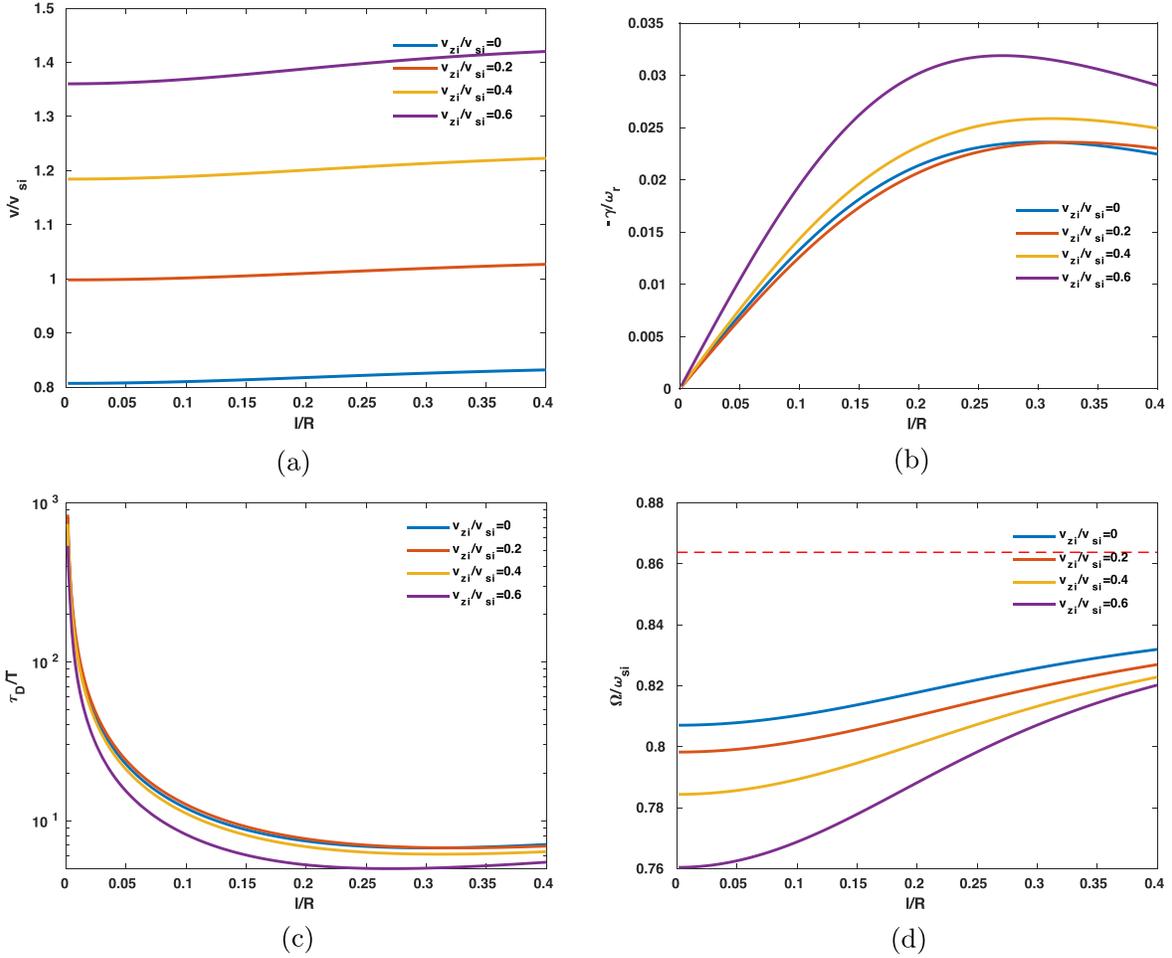

**Figure 5.** (a) The phase speed $v/v_{si} \equiv \omega_r/\omega_{si}$, (b) the damping rate to frequency ratio $-\gamma/\omega_r$, (c) the damping time to period ratio $\tau_D/T = \omega/(2\pi|\gamma|)$, and (d) the Doppler shifted phase speed $\Omega/\omega_{si}$ as a function of $l/R$ for the slow surface sausage modes for $k_z R = 0.7$, $B_{\phi_i}/B_{zi} = 0.5$, and different flow parameters. Other parameters of the tube are the same as in Figure 3.

Figure 4 shows variations of the phase speed (or normalized frequency) $v/v_{si} \equiv \omega_r/\omega_{si}$, the damping rate to frequency ratio $-\gamma/\omega_r$, the damping time to period ratio $\tau_D/T = \omega/(2\pi|\gamma|)$ and Doppler shifted phase speed $\Omega/\omega_{si}$ of the slow surface sausage modes versus $k_z R$ for various values of the twist parameter $B_{\phi_i}/B_{zi} = (0, 0.2, 0.4, 0.6)$ and for the thickness of the inhomogeneous layer $l/R = 0.1$ and the flow parameter $v_{zi}/v_{si} = 0.6$. This figure shows that (i) the minimum value of the phase speed $v/v_{si}$ decreases and shifts to smaller $k_z R$ with increasing the twist parameter. (ii) When the twist parameter increases, the maximum value of $-\gamma/\omega_r$ increases and shifts to smaller $k_z R$. (iii) The minimum value of the damping time to period ratio decreases when the twist parameter increases, and our numerical calculations show that the reduction is larger for static tubes. For example, the minimum value of $\tau_D/T$ for $v_{zi}/v_{si} = 0.6$ and $B_{\phi_i}/B_{zi} = 0.4$ decreases $\sim 23\%$ but for $v_{zi}/v_{si} = 0$ and $B_{\phi_i}/B_{zi} = 0.4$ it decreases by $\sim 49\%$ compared to the case of the untwisted tubes.

In Figures 5 and 6 we have plotted the variation of $v/v_{si}$, $-\gamma/\omega_r$, $\tau_D/T$, and $\Omega/\omega_{si}$ versus the thickness of the inhomogeneous layer $l/R$ for different flow parameters $v_{zi}/v_{si} = (0, 0.2, 0.4, 0.6)$ with magnetic twist $B_{\phi_i}/B_{zi} = 0.5$ and $k_z R = (0.7, 2)$. This figure shows that (i) for a given flow parameter $v_{zi}/v_{si}$, the value of phase speed $v/v_{si}$ and the Doppler shifted phase speed $\Omega/\omega_{si}$ increase with increasing $l/R$. (ii) With increasing $l/R$ the ratio of the damping rate to oscillation frequency $-\gamma/\omega_r$ reaches a maximum value and then decreases. (iii) The maximum value of $-\gamma/\omega_r$ increases and moves to smaller $l/R$ when $v_{zi}/v_{si}$ increases. For example, the value of $-\gamma/\omega_r$ when the flow parameter $v_{zi}/v_{si}$ increases from 0 to 0.6 for the case $B_{\phi_i}/B_{zi} = 0.5$ and $k_z R = 0.7$ increases $\sim 26\%$ and for the case $B_{\phi_i}/B_{zi} = 0.5$ and $k_z R = 2$ increases $\sim 71\%$ compared to the case of no twist.

Panels (a) and (b) of Figure 7 show the minimum value of the damping time to period ratio $\tau_D/T = \omega/(2\pi|\gamma|)$ as a function of the flow speed $v_{zi}/v_{si}$ and twist parameter $B_{\phi_i}/B_{zi}$, respectively, where the thickness of the inhomogeneous layer is $l/R = 0.1$. Panel (a) shows the results for different twist parameters $B_{\phi_i}/B_{zi} = (10^{-5}, 0.3, 0.5)$. This panel indicates that for small $v_{zi}/v_{si}$ the effect of the twist is important but for the larger flow parameter ($v_{zi}/v_{si} > 0.6$) the effect of the magnetic twist is negligible. Panel (b) represents $\tau_D/T = \omega/(2\pi|\gamma|)$ as a function of the twist parameter for various values of flow parameter $v_{zi}/v_{si} = (0, 0.2, 0.4, 0.6, 0.8, 1)$. This panel shows that for smaller $B_{\phi_i}/B_{zi}$ the effect of the flow parameter is much more important but for larger $B_{\phi_i}/B_{zi}$ the effect of flow is less important. It should be noted that to plot Figure 7, the minimum value of the damping time to period ratio in the observational

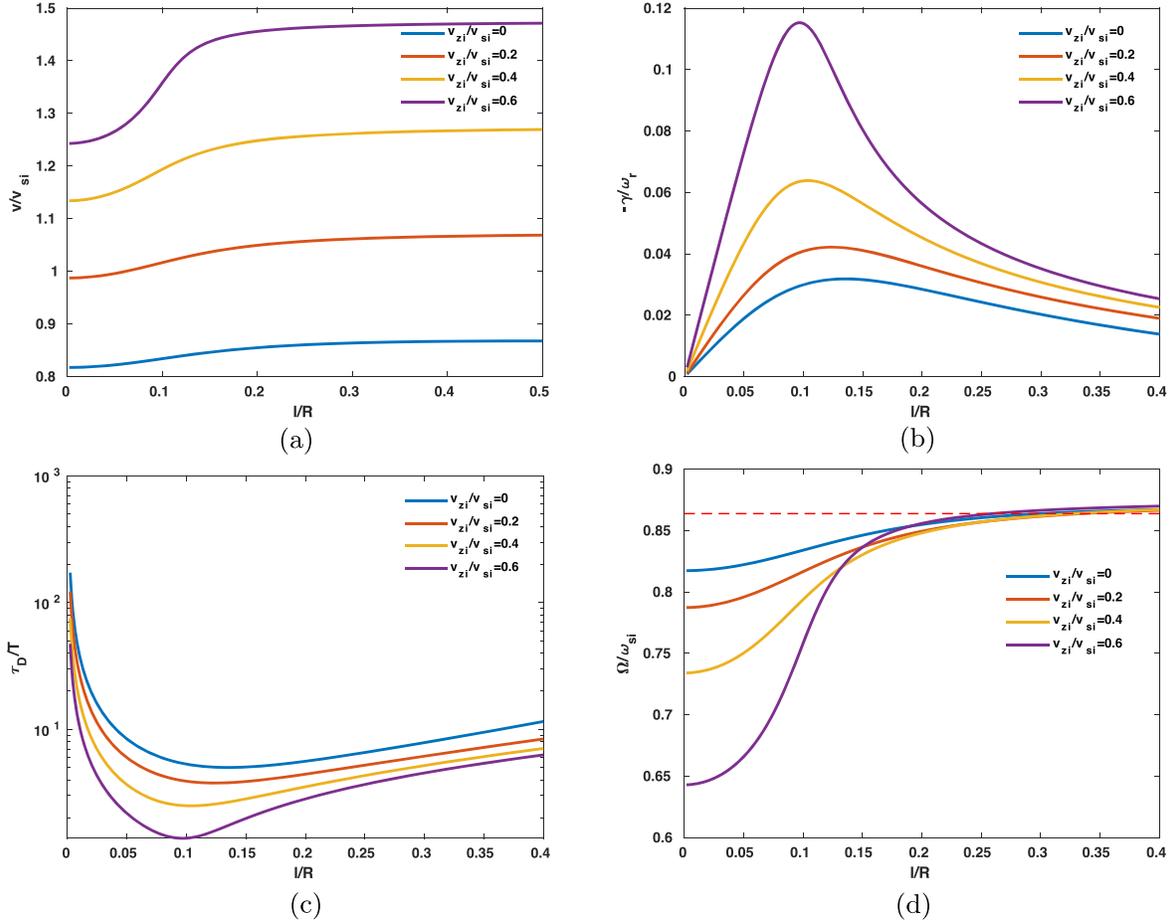

**Figure 6.** Same as Figure 5, but for $k_z R = 2$.

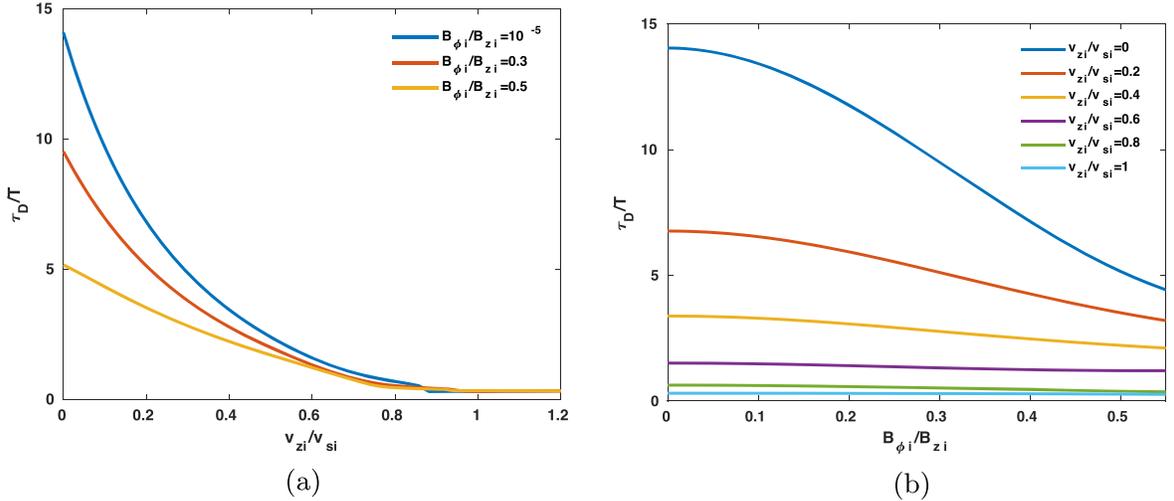

**Figure 7.** The minimum value of the damping time to period ratio ($\tau_D/T$) for the slow surface sausage modes vs. (a) the flow speed ($v_{zi}/v_{si}$) and (b) the magnetic twist ($B_{\phi i}/B_{zi}$) for $l/R = 0.1$. Here, we have $k_z R \leqslant 5$ and the other parameters are the same as in Figure 3.

wavelength region, i.e., $k_z R \leqslant 5$, is used. In the presence of both the plasma flow and magnetic twist the ratio of the damping time to the oscillation period can reach about 0.27, which is not much different from the result obtained by Sadeghi et al. (2021).

## 5. Conclusions

In this paper, the effect of both magnetic twist and plasma flow on slow sausage waves in magnetic flux tubes under solar photospheric (or magnetic pore) conditions has been investigated. We considered a cylindrical flux tube in which the density, longitudinal magnetic field, and flow plasma are constant inside and outside of the flux tube but the density and the longitudinal and azimuthal components of the magnetic field and the plasma flow change continuously in the inhomogeneous layer. Using the solutions of the MHD equations in the core and external regions and the connection formula, we obtained the dispersion relation (42) and solved it numerically, and calculated the phase speed (or normalized frequency) $v/v_{si} \equiv \omega_r/\omega_{si}$, the damping rate to frequency ratio $-\gamma/\omega_r$, the damping time to period ratio $\tau_D/T = \omega_r/(2\pi|\gamma|)$, and the Doppler shifted frequency $\Omega/\omega_{si}$ of the slow surface sausage modes under the magnetic pore conditions. The dispersion relation has been solved numerically. The numerical results show that:

1. For given values of the thickness of the inhomogeneous layer $l/R$, and the twist parameter $B_{\phi_i}/B_{zi}$, as the flow parameter $v_{zi}/v_{si}$ increases, (i) the minimum values of both the Doppler shifted frequency $\Omega/\omega_{si}$ and the damping time to period ratio $\tau_D/T = \omega_r/(2\pi|\gamma|)$ decrease, and (ii) the maximum value of $-\gamma/\omega_r$ increases.
2. The effect of plasma flow in decreasing the ratio of the damping time to the oscillation period is more strong in untwisted tubes, i.e., as the twist parameter increases the effect of plasma flow becomes less important.
3. The effect of magnetic twist in decreasing the ratio of the damping time to the oscillation period is more effective in flux tubes with a small flow parameter. For example, in flux tubes with flow parameter $v_{zi}/v_{si} = 0.2$, when the twist parameter $B_{\phi_i}/B_{zi}$ increases from 0.1 to 0.3 the minimum value of $\tau_D/T$ decreases $\sim 49\%$ with respect to untwisted tubes. But in flux tubes with flow parameter $v_{zi}/v_{si} = 0.6$, when the twist parameter $B_{\phi_i}/B_{zi}$ increases from 0.1 to 0.3 the minimum value of $\tau_D/T$ decreases by $\sim 17\%$ with respect to untwisted tubes.
4. For the case of $l/R = 0.1$, the minimum value of $\tau_D/T$ for $B_{\phi_i}/B_{zi} = 0.3$, for instance, decreases $\sim 97\%$ when the flow parameter $v_{zi}/v_{si}$ changes from 0 to 1. This value for $B_{\phi_i}/B_{zi} = 0$, decreases $\sim 98\%$ when the flow parameter $v_{zi}/v_{si}$ changes from 0 to 1.

The authors thank the anonymous referee for their valuable comments.

## Appendix A
## Constants of Background Gas Pressure

Here, the parameters used in Equation (10) governing the radial dependence of background pressure are given by

$$A_1 = p_0 - \left(1 - \frac{\mu_0 \rho_i A_V^2}{2A^2}\right) A^2 r_i^2/\mu_0 - A_2 r_i - A_3 r_i^2 - A_4 r_i^3,$$

$$A_2 \equiv -\frac{3A^2 r_i^2(\zeta_{B_\phi}^2 - 1) + B_{zi}^2(\zeta_{B_z}^2 - 1)}{2\mu_0 l} + \frac{\rho_i A_V^2 r_i^2}{l^3(\zeta_{B_z}^2 - 1)^3}(-l(\zeta_{V_z} - 1)(\zeta_{B_z}^2 - 1)(2(-l)(\zeta_{B_\phi}^2 - 1)(\zeta_{B_z}^2 - 1)$$
$$+ (\zeta_{V_z} - 1)(-l(\zeta_{B_z}^2 - 1) + (\zeta_{B_\phi}^2 - 1)(l + 2r_i(\zeta_{B_z}^2 - 1))))$$
$$+(\zeta_\rho - 1)(l^2(\zeta_{B_\phi}^2 - 1)(\zeta_{B_z}^2 - 1)^2 + 2l(\zeta_{V_z} - 1)(\zeta_{B_z}^2 - 1)(l(\zeta_{B_z}^2 - 1) + (\zeta_{B_\phi}^2 - 1)(-l - 2r_i(\zeta_{B_z}^2 - 1)))$$
$$+(\zeta_{V_z} - 1)^2(-l(\zeta_{B_z}^2 - 1)(l + 2r_i(\zeta_{B_z}^2 - 1)) + (\zeta_{B_\phi}^2 - 1)(l^2 + 2lr_i(\zeta_{B_z}^2 - 1) + 3r_i^2(\zeta_{B_z}^2 - 1)^2)))),$$

$$A_3 \equiv \frac{\rho_i A_V^2 r_i^2}{2l^3(\zeta_{B_z}^2 - 1)^2}(\zeta_{V_z} - 1)(l(\zeta_{V_z} - 1)(\zeta_{B_\phi}^2 - 1)(\zeta_{B_z}^2 - 1)$$
$$+(\zeta_\rho - 1)(2l(\zeta_{B_\phi}^2 - 1)(\zeta_{B_z}^2 - 1) + (\zeta_{V_z} - 1)(l(\zeta_{B_z}^2 - 1) + (\zeta_{B_\phi}^2 - 1)(-l - 3r_i(\zeta_{B_z}^2 - 1))))),$$

$$A_4 \equiv \frac{\rho_i A_V^2 r_i^2 (\zeta_\rho - 1)(\zeta_{V_z} - 1)^2(\zeta_{B_\phi}^2 - 1)}{3l^3(\zeta_{B_z}^2 - 1)},$$

$$A_5 \equiv -A^2 r_i^2/\mu_0 \left(1 - \frac{r_i(\zeta_{B_\phi}^2 - 1)}{(r_e - r_i)}\right) + \frac{\rho_i A_V^2 r_i^2}{l^3(l - r_i(\zeta_{B_z}^2 - 1))}(l - r_i(\zeta_\rho - 1))(-l + r_i(\zeta_{V_z} - 1))^2(l - r_i(\zeta_{B_\phi}^2 - 1)),$$

$$A_6 \equiv \rho_i A_V^2 r_i^2 \frac{l((\zeta_\rho - 1) - (\zeta_{B_z}^2 - 1))((\zeta_{V_z} - 1) - (\zeta_{B_z}^2 - 1))^2(-(\zeta_{B_\phi}^2 - 1) + (\zeta_{B_z}^2 - 1))}{(\zeta_{B_z}^2 - 1)^4(-l + r_i(\zeta_{B_z}^2 - 1))},$$

$$p_1 = p_0 - \left(1 - \frac{\mu_0 \rho_i A_V^2}{2A^2}\right) A^2 r_i^2/\mu_0 + A_2(r_e - r_i) + A_3(r_e^2 - r_i^2) + A_4(r_e^3 - r_i^3) + A_5 \ln(r_e/r_i) + A_6 \ln(1 + (\zeta_{B_z}^2 - 1)). \quad (A1)$$

The continuity of the total pressure at the boundaries of the tube, i.e., at the inner ($r = r_i$) and outer ($r = r_e$) radii of the tube has been used to determine $A_1$ and $p_1$ in terms of other constants. In the case of photospheric conditions, these parameters can be simplified as

$$A_1 = p_0 - \left(1 - \frac{\mu_0 \rho_i A_V^2}{2A^2}\right) A^2 r_i^2 / \mu_0 - A_2 r_i - A_3 r_i^2 - A_4 r_i^3,$$

$$A_2 \equiv \frac{3A^2 r_i^2 + B_0^2}{2\mu_0 l} - \frac{\rho_i A_V^2 r_i^2}{l^2}[(2r_i - 2l) - (\zeta_\rho - 1)(l + 4r_i) + 3r_i^2/l],$$

$$A_3 \equiv \frac{\rho_i A_V^2 r_i^2}{2l^3}(l + (\zeta_\rho - 1)(3r_i - 2l)),$$

$$A_4 \equiv \frac{\rho_i A_V^2 r_i^2 (\zeta_\rho - 1)}{3l^3},$$

$$A_5 \equiv -A^2 r_i^2 / \mu_0 \left(1 + \frac{r_i}{(r_e - r_i)}\right) + \frac{\rho_i A_V^2 r_i^2 r_e}{l^3}(l - r_i(\zeta_\rho - 1)),$$

$$A_6 = 0,$$

$$p_1 = p_0 - \left(1 - \frac{\mu_0 \rho_i A_V^2}{2A^2}\right) A^2 r_i^2 / \mu_0 + A_2(r_e - r_i) + A_3(r_e^2 - r_i^2) + A_4(r_e^3 - r_i^3) + A_5 \ln(r_e/r_i). \quad (A2)$$

## Appendix B
## Resonant Position

Equation (50) can be rewritten as a fourth-order equation in terms of $\delta_c$ as follows:

$$\mathcal{A}\delta_c^4 + \mathcal{B}\delta_c^3 + \mathcal{C}\delta_c^2 + \mathcal{E}\delta_c + \mathcal{F} = 0. \quad (B1)$$

The constants $\mathcal{A}, \mathcal{B}, \mathcal{C}, \mathcal{E}$, and $\mathcal{F}$ are given by

$$\mathcal{A} \equiv \frac{\gamma A_4 l^3}{v_{Ai}^2 v_{si}^2}(v_c^2(\zeta_\rho - 1) - (\zeta_\rho v_{Ae}^2 - 1)), \quad (B2)$$

$$\mathcal{B} \equiv \frac{\gamma A_4 l^3}{v_{Ai}^2 v_{si}^2}(v_c^2 - 1) + \frac{\gamma}{2v_{Ai}^2 v_{si}^2}(v_c^2(\zeta_\rho - 1) - (\zeta_\rho v_{Ae}^2 - 1))\left(2A_3 l^2 + 6A_4 l^2 r_i - A_5\left(\frac{l}{r_i}\right)^2 - A_6(\zeta_{B_z} - 1)^2\right), \quad (B3)$$

$$\mathcal{C} \equiv 1 + \frac{v_c^2}{v_{ci}^2}(\zeta_\rho - 1) + \zeta_\rho\left[\frac{2v_c^2}{v_{ci}^2} - (v_{sei}^2 + v_{Aei}^2)\right] - \zeta_\rho^2\left(\frac{v_c^2}{v_{ci}^2} - v_{sei}^2 v_{Aei}^2\right)$$
$$+ \frac{\gamma}{2v_{Ai}^2 v_{si}^2}\left(\frac{1}{v_{si}} - 1\right)\left(2A_3 l^2 + 6A_4 l^2 r_i - A_5\left(\frac{l}{r_i}\right)^2\right)$$
$$+ \frac{\gamma}{v_{Ai}^2 v_{si}^2}(v_c^2(\zeta_\rho - 1) - (\zeta_\rho v_{Ae}^2 - 1))\left(A_3(2lr_i - r_e^2) + A_4(3lr_i^2 - r_e^3) + A_5\left(\frac{l}{r_i} - \ln\left(\frac{r_e}{r_i}\right)\right)\right), \quad (B4)$$

$$\mathcal{E} \equiv 2\left(\frac{v_c^2}{v_{ci}^2} - 1\right) - \chi\left[\frac{v_c^2}{v_{ci}^2}\left(1 + \frac{v_{se}^2 + v_{Ae}^2}{v_{si}^2 + v_{Ai}^2}\right) - (v_{sei}^2 + v_{Aei}^2)\right]$$
$$+ \frac{\gamma}{v_{Ai}^2 v_{si}^2}\left(\frac{1}{v_{si}} - 1\right)\left(A_3(2lr_i - r_e^2) + A_4(3lr_i^2 - r_e^3) + A_5\left(\frac{l}{r_i} - \ln\left(\frac{r_e}{r_i}\right)\right)\right)$$
$$+ \frac{\gamma}{v_{Ai}^2 v_{si}^2}(v_c^2(\zeta_\rho - 1) - (\zeta_\rho v_{Ae}^2 - 1))(A_3 r_i^2 + A_4 r_i^3), \quad (B5)$$

$$\mathcal{F} \equiv \frac{v_c^2}{v_{ci}^2} - 1 + \frac{\gamma r_i^2}{v_{Ai}^2 v_{si}^2}\left(\frac{1}{v_{si}} - 1\right), \quad (B6)$$

where we have used the following approximation

$$\ln(r/r_i) = \ln\left[1 + \left(\frac{r_e - r_i}{r_i}\right)\delta_c\right],$$
$$\simeq \left(\frac{r_e - r_i}{r_i}\right)\delta_c - \frac{1}{2}\left(\frac{r_e - r_i}{r_i}\right)^2 \delta_c^2, \tag{B7}$$

for $\left(\frac{r_e - r_i}{r_i}\right)\delta_c < 1$ and considered only the terms up to $O(\delta_c^2)$. It has been checked that keeping the higher order terms $O(\delta_c^3)$ does not affect the results. This fourth-order equation has been used to obtain the cusp resonance absorption point ($r_c$) in numerical calculations.

## Appendix C
## Weak Damping Rate for the Surface Sausage Mode

Here, Equation (54) has been used to obtain an analytical expression for the damping rate of slow surface sausage waves in the weak damping limit, i.e., $\gamma \ll \omega_r$. To do this, we first calculate $\frac{\partial D_{AR}}{\partial \omega}$ from Equation (43) as follows:

$$\frac{\partial D_{AR}}{\partial \omega} = 2\rho_i \omega - 2\rho_e \omega \frac{k_{ri}}{k_{re}} \mathcal{Q} - \rho_e(\omega - \omega_{Ae}^2)\left(\frac{1}{k_{re}}\frac{dk_{ri}}{d\omega} - \frac{k_{ri}}{k_{re}^2}\frac{dk_{re}}{d\omega}\right)\mathcal{Q} - \rho_e(\omega^2 - \omega_{Ae}^2)\frac{k_{ri}}{k_{re}}\frac{d\mathcal{Q}}{d\omega}. \tag{C1}$$

Now from Equation (36), one can obtain

$$\frac{dk_{ri}}{dw} = \frac{-\omega^3(\omega^2 - 2\omega_{ci}^2)}{(v_{si}^2 + v_{Ai}^2)(\omega^2 - \omega_{ci}^2)^2 k_{ri}}, \tag{C2}$$

$$\frac{dk_{re}}{dw} = \frac{-\omega^3(\omega^2 - 2\omega_{ce}^2)}{(v_{se}^2 + v_{Ae}^2)(\omega^2 - \omega_{ce}^2)^2 k_{re}}. \tag{C3}$$

Also from Equation (45) for $\frac{d\mathcal{Q}}{d\omega}$, we get

$$\frac{d\mathcal{Q}}{d\omega} = \frac{\mathcal{Q}}{x}\frac{dx}{dw} - x\frac{\frac{d}{d\omega}\left(\frac{\mu_0 D_e(1-\nu) - 2B_{\phi e}^2 n_e^2}{\mu_0 D_e y} - \frac{K_{\nu-1}(y)}{K_\nu(y)}\right)}{\left[\frac{n_i + k_z}{k_z}s - 2\frac{M(a,b-1;s)}{M(a,b;s)}\right]}$$
$$+ x\frac{\left(\frac{\mu_0 D_e(1-\nu) - 2B_{\phi e}^2 n_e^2}{\mu_0 D_e y} - \frac{K_{\nu-1}(y)}{K_\nu(y)}\right)\frac{d}{d\omega}\left[\frac{n_i + k_z}{k_z}s - 2\frac{M(a,b-1;s)}{M(a,b;s)}\right]}{\left[\frac{n_i + k_z}{k_z}s - 2\frac{M(a,b-1;s)}{M(a,b;s)}\right]^2}. \tag{C4}$$

After some algebra, this equation can be written as

$$\frac{d\mathcal{Q}}{d\omega} = \frac{\mathcal{Q}}{x}\frac{dx}{dw} + x\frac{\left(\frac{1-\nu}{y^2} + \frac{2v_{A\phi_e}^2 n_e^2}{(\omega^2 - \omega_{Ae}^2)y^2} + \left(\frac{K_{\nu-1}'}{K_\nu} - \frac{K_\nu' K_{\nu-1}}{K_\nu^2}\right)\right)\frac{dy}{d\omega} + \frac{1}{y}\frac{d\nu}{d\omega} - \frac{4v_{A\phi_e}^2}{(\omega^2 - \omega_{Ae}^2)y}\left(n_e \frac{dn_e}{d\omega} - \frac{\omega n_e^2}{(\omega^2 - \omega_{Ae}^2)}\right)}{\left[\frac{n_i + k_z}{k_z}s - 2\frac{M(a,b-1;s)}{M(a,b;s)}\right]}$$
$$+ x\frac{\left[\frac{s}{k_z}\frac{dn_i}{d\omega} + \frac{n_i + k_z}{k_z}ds - 2\frac{d}{d\omega}\left(\frac{M(a,b-1;s)}{M(a,b;s)}\right)\right]\left(\frac{\mu_0 D_e(1-\nu) - 2B_{\phi e}^2 n_e^2}{\mu_0 D_e y} - \frac{K_{\nu-1}(y)}{K_\nu(y)}\right)}{\left[\frac{n_i + k_z}{k_z}s - 2\frac{M(a,b-1;s)}{M(a,b;s)}\right]^2}, \tag{C5}$$

where $x = k_{ri} r_i$ and $y = k_{re} r_e$. In terms of new functions, this simplifies as

$$\frac{d\mathcal{Q}}{d\omega} = \mathcal{P}\frac{dx}{dw} + \mathcal{S}\frac{dy}{d\omega}, \tag{C6}$$

where

$$\mathcal{P} = \frac{\mathcal{Q}}{x} + x \frac{\left[\frac{s}{k_z}\frac{dn_i}{d\omega} + \frac{n_i + k_z}{k_z}ds - 2\frac{d}{d\omega}\left(\frac{M(a,b-1;s)}{M(a,b;s)}\right)\right]\left(\frac{\mu_0 D_e(1-\nu) - 2B_{\phi e}^2 n_e^2}{\mu_0 D_e y} - \frac{K_{\nu-1}(y)}{K_\nu(y)}\right)}{\frac{dx}{dw}\left[\frac{n_i + k_z}{k_z}s - 2\frac{M(a,b-1;s)}{M(a,b;s)}\right]^2}, \tag{C7}$$

$$\mathcal{S} = x \frac{\left(\frac{1-\nu}{y^2} + \frac{2v_{A\phi_e}^2 n_e^2}{(\omega^2 - \omega_{Ae}^2)y^2} + \left(\frac{K_{\nu-1}'}{K_\nu} - \frac{K_\nu' K_{\nu-1}}{K_\nu^2}\right)\right)\frac{dy}{d\omega} + \frac{1}{y}\frac{d\nu}{d\omega} - \frac{4v_{A\phi_e}^2}{(\omega^2 - \omega_{Ae}^2)y}\left(n_e\frac{dn_e}{d\omega} - \frac{\omega n_e^2}{(\omega^2 - \omega_{Ae}^2)}\right)}{\frac{dy}{d\omega}\left[\frac{n_i + k_z}{k_z}s - 2\frac{M(a,b-1;s)}{M(a,b;s)}\right]}. \tag{C8}$$

In addition, from Equation (22) we have

$$\frac{dn_i}{dw} = \frac{\omega^3(\omega^2 - 2\omega_{ci}^2)}{(v_{si}^2 + v_{Ai}^2)(\omega^2 - \omega_{ci}^2)^2 n_i}, \tag{C9}$$

$$\frac{dn_e}{dw} = \frac{\omega^3(\omega^2 - 2\omega_{ce}^2)}{(v_{se}^2 + v_{Ae}^2)(\omega^2 - \omega_{ce}^2)^2 n_e}. \tag{C10}$$

With the help of Equations (C2) and (C3), Equation (C6) takes the form

$$\frac{d\mathcal{Q}}{d\omega} = x\mathcal{P}\frac{\omega^3(\omega^2 - 2\omega_{ci}^2)}{(\omega_{si}^2 - \omega^2)(\omega_{Ai}^2 - \omega^2)(\omega^2 - \omega_{ci}^2)} + y\mathcal{S}\frac{\omega^3(\omega^2 - 2\omega_{ce}^2)}{(\omega_{se}^2 - \omega^2)(\omega_{Ae}^2 - \omega^2)(\omega^2 - \omega_{ce}^2)}. \tag{C11}$$

Substituting this into Equation (C1) yields

$$\frac{\partial D_{AR}}{\partial \omega} = 2\rho_i\omega - 2\rho_e\omega\frac{k_{ri}}{k_{re}}\mathcal{Q} - \rho_e\omega^3(\omega^2 - \omega_{Ae}^2)\frac{k_{ri}}{k_{re}}$$
$$\times \left(\frac{(\mathcal{Q} + x\mathcal{P})(\omega^2 - 2\omega_{ci}^2)}{(\omega_{si}^2 - \omega^2)(\omega_{Ai}^2 - \omega^2)(\omega^2 - \omega_{ci}^2)} - \frac{(\mathcal{Q} - y\mathcal{S})(\omega^2 - 2\omega_{ce}^2)}{(\omega_{se}^2 - \omega^2)(\omega_{Ae}^2 - \omega^2)(\omega^2 - \omega_{ce}^2)}\right). \tag{C12}$$

Finally, using Equations (44) and (C12) in Equation (54) one can obtain the damping rate $\gamma$ in the limit of weak damping for the slow surface sausage modes in the cusp continuum as

$$\gamma|_{\omega=\omega_r} = -\frac{\frac{\pi\rho_e k_z^2}{k_{re}\rho_c|\Delta_c|}\bigg|_{r=r_c}\left(\frac{v_s^2}{v_A^2 + v_s^2}\right)^2((\omega^2 - \omega_{Ai}^2) + \mathcal{Z})(\omega^2 - \omega_{Ae}^2)\mathcal{G}}{2\omega\left(1 - \chi\frac{k_{ri}}{k_{re}}Q\right) - \omega\chi\mathcal{T}}, \tag{C13}$$

where

$$\mathcal{T} = \omega_r^2(\omega_r^2 - \omega_{Ae}^2)\frac{k_{ri}}{k_{re}}\left(\frac{(\mathcal{Q} + x\mathcal{P})(\omega_r^2 - 2\omega_{ci}^2)}{(\omega_{si}^2 - \omega_r^2)(\omega_{Ai}^2 - \omega_r^2)(\omega_r^2 - \omega_{ci}^2)} - \frac{(\mathcal{Q} - y\mathcal{S})(\omega_r^2 - 2\omega_{ce}^2)}{(\omega_{se}^2 - \omega_r^2)(\omega_{Ae}^2 - \omega_r^2)(\omega_r^2 - \omega_{ce}^2)}\right). \tag{C14}$$

## ORCID iDs

Mohammad Sadeghi  https://orcid.org/0000-0002-6409-7679
Karam Bahari  https://orcid.org/0000-0002-4838-5646
Kayoomars Karami  https://orcid.org/0000-0003-0008-0090